\documentclass[12pt, a4paper]{article}

\usepackage{amssymb}
\usepackage{amsmath}
\usepackage{amsfonts}
\usepackage{graphicx}
\usepackage{mathrsfs}
\usepackage{comment}
\usepackage{cite}
\def\->{\rightarrow}
\def\<-{\leftarrow}

\newcommand{\Slash}[1]{{\ooalign{\hfil#1\hfil\crcr\raise.167ex\hbox{/}}}}

\newcommand{\beq}{\begin{equation}}  \newcommand{\eeq}{\end{equation}}
\newcommand{\bef}{\begin{figure}}  \newcommand{\eef}{\end{figure}}
\newcommand{\bec}{\begin{center}}  \newcommand{\eec}{\end{center}}
  
\newcommand{\laq}[1]{\label{eq:#1}}  

\newcommand{\Eq}[1]{Eq.~(\ref{eq:#1})}

\newcommand{\eq}[1]{(\ref{eq:#1})}
\newcommand{\Sec}[1]{Sec.~\ref{chap:#1}}
\newcommand{\ab}[1]{\left|{#1}\right|}

\newcommand{\lac}[1]{\label{chap:#1}}

\newcommand{\bed}{\begin{description} \item}
\newcommand{\eed}{\end{description}}

\def\({\left(}
\def\){\right)}

\def\O{\mathcal{O}}
\def\U{\mathop{\rm U}}

\def\a{\alpha}

\def\d{\delta}

\def\f{\phi}
\def\g{\gamma}
\def\h{\theta}

\def\l{\lambda}
\def\m{\mu}

\def\p{\psi}

\def\r{\rho}
\def\s{\sigma}
\def\t{\tau}

\def\x{\xi}

\def\D{\Delta}
\def\G{\Gamma}

\def\L{\Lambda}

\def\ol{\overline}
\def\tl{\tilde}
\def\*{\dagger}

\newcommand{\AND}{~{\rm and}~}
\newcommand{\EV}{ {\rm eV} }

\newcommand{\GEV}{ {\rm\, GeV} }
\newcommand{\TEV}{ {\rm TeV} }

\renewcommand{\thefootnote}{\fnsymbol{footnote}}

\begin{document}
\begin{titlepage}
\begin{center}

\hfill TU-1049,  IPMU17-0145\\

\vskip .75in

{\Large\bf The ALP miracle revisited
}

\vskip .75in

{ \large Ryuji Daido\,$^{a}$\footnote{email: daido@tuhep.phys.tohoku.ac.jp},
    Fuminobu  Takahashi\,$^{a,b}$\footnote{email: fumi@tuhep.phys.tohoku.ac.jp},
    Wen Yin\,$^{c}$\footnote{email: wyin@ihep.ac.cn}}

\vskip 0.25in

\begin{tabular}{ll}
$^{a}$ &\!\! {\em Department of Physics, Tohoku University, }\\
& {\em Sendai, Miyagi 980-8578, Japan}\\[.3em]
$^{b}$ &\!\! {\em Kavli IPMU (WPI), UTIAS,}\\
&{\em The University of Tokyo,  Kashiwa, Chiba 277-8583, Japan}\\[.3em]
$^{c}$ &\!\! {\em Institute of High Energy Physics,}\\
&{\em  Chinese Academy of Sciences, Beijing 100049,  China}\\[.3em]
& {\em }

\end{tabular}

\vskip .5in

\begin{abstract}
We revisit the ALP miracle scenario where the inflaton and dark matter are unified by a single axion-like particle (ALP). 
We first extend our previous analysis on the inflaton dynamics to identify the whole viable parameter space consistent with 
the CMB observation. Then, we evaluate the relic density of the ALP dark matter by incorporating uncertainties of the 
model-dependent couplings to the weak gauge bosons as well as the dissipation effect.
The preferred ranges of  the ALP mass and coupling to photons are found to be
$0.01\lesssim m_\phi  \lesssim 1$\,eV and $g_{\phi \gamma \gamma} = {\cal O}(10^{-11})$\,GeV$^{-1}$,
which slightly depend on these uncertainties.
 Interestingly, the preferred regions are within reach of future solar axion helioscope experiments, IAXO and TASTE, and laser-based stimulated photon-photon collider experiments.  We also discuss possible extensions of the ALP miracle scenario by introducing interactions of the ALP with
fermions. 
\end{abstract}

\end{center}
\end{titlepage}
\setcounter{footnote}{0}
\setcounter{page}{1}

\renewcommand{\thefootnote}{\arabic{footnote}}

\section{Introduction}
The observations of temperature and polarization anisotropies of cosmic microwave background radiation (CMB) showed that our Universe experienced exponential expansion called inflation at a very early stage of the evolution and that the present Universe is dominated by dark matter and dark energy~\cite{Ade:2015lrj}. In particular, inflaton and dark matter are clear evidence for physics beyond the standard model (SM), and a variety of models and their experimental and observational implications have been studied extensively. 

The inflaton and dark matter have something in common; both are electrically neutral and occupied a significant fraction of the Universe
at different times.  The unification of inflaton and dark matter has been discussed in the literature, where the reheating is incomplete, and the remaining relic inflatons become dark matter at later times~\cite{Kofman:1994rk,Kofman:1997yn,Mukaida:2014kpa,Bastero-Gil:2015lga}\footnote{
Thermalized inflaton particles~\cite{Lerner:2009xg,Okada:2010jd,Khoze:2013uia,Bastero-Gil:2015lga} or inflatino~\cite{Nakayama:2010kt} 
can also be a WIMP dark matter.}.
One of the most non-trivial requirements for such models is to stabilize the inflaton in the present vacuum. To this end, the inflaton is often assumed to be charged under $Z_2$ symmetry~\cite{Kofman:1994rk,Kofman:1997yn,Mukaida:2014kpa} or to be coupled to other fields in such a way that the decay occurs only when the inflaton oscillation amplitude is sufficiently large~\cite{Bastero-Gil:2015lga}, etc.

Recently, the present authors proposed a novel way to realize the unification of inflaton and dark matter by a single axion-like particle (ALP)~\cite{Daido:2017wwb}.  In this scenario, the inflaton is an ALP with a sub-Planckian decay constant, and the 
potential consists of two cosine functions which conspire to make the curvature around the top of the potential sufficiently small for the slow-roll inflation to take place. This is the minimal realization of the so-called multi-natural inflation~\cite{Czerny:2014wza,Czerny:2014xja,Czerny:2014qqa,Higaki:2014sja}. The reason why we need multiple cosine functions is that, with a single cosine function, the inflaton field excursion exceeds the Planck scale which would necessitate a super-Planckian decay constant~\cite{Freese:1990rb,Adams:1992bn}. The super-Planckian decay constant is often questioned by the argument based on the quantum gravity (e.g. see Refs.~\cite{Rudelius:2014wla,delaFuente:2014aca,
Rudelius:2015xta,Montero:2015ofa,Brown:2015iha}), and moreover, even if the super-Planckian decay constant is somehow justified, the predicted spectral index and tensor-to-scalar ratio are not preferred by the current CMB data~\cite{Ade:2015lrj}.  

The key feature of our scenario is that the inflaton masses at the potential maximum and minimum have the same absolute magnitude but a different sign. In other words, the slow-roll condition for the inflaton implies that the inflaton is light also at the potential minimum. Therefore, the longevity of dark matter is a direct consequence of the slow-roll inflation. For successful reheating we introduced a coupling of the inflaton (ALP) to photons, and showed that successful reheating takes place for the axion-photon coupling, $g_{\phi \gamma \gamma} \gtrsim {\cal O}(10^{-11})$\,GeV$^{-1}$ (see \Eq{int} for the definition). 
In contrast to the conventional scenario, the dissipation process plays the major role in reheating. 
Since the dissipation rate decreases with the oscillation amplitude, a small amount of the ALP condensate remains and contributes to dark matter. 
Interestingly,  the relic ALP condensate can explain the observed dark matter abundance if 
$m_\f =\O(0.01-0.1)\,\EV \AND g_{\phi \gamma \gamma} = {\cal O}(10^{-11})$\,GeV$^{-1}$, which is within the reach of the future solar 
axion helioscope experiments, 
IAXO~\cite{Irastorza:2011gs,Armengaud:2014gea} and TASTE~\cite{Anastassopoulos:2017kag}, and laser-based photon colliders~\cite{Hasebe:2015jxa,Fujii:2010is,Homma:2017cpa}.\footnote{ Recently an ALP coupling to photon is a hot experimental target and many experiments are proposed to search for it, see e.g. \cite{Alekhin:2015byh, Capparelli:2015mxa, Dobrich:2015jyk, Brubaker:2016ktl, Spector:2016vwo, TheMADMAXWorkingGroup:2016hpc, Stern:2016bbw, Woohyun:2016hkn, Choi:2017hjy, Inada:2017lop, Budnev:2017jyl, Alesini:2017ifp}.}  
We call such a coincidence as the ALP miracle~\cite{Daido:2017wwb}. 

In this paper, we revisit the axionic unification of the inflaton and dark matter to study in detail the inflaton dynamics as well as the reheating process by scanning a whole parameter space. In particular, only a part of the inflation model parameters was explored in Ref.~\cite{Daido:2017wwb}, and
we made an only order-of-magnitude estimate of the plausible ALP mass and coupling to photons. In this paper, we evaluate the ALP parameters more precisely to see to what extent the future experiments will be able to cover the predicted parameter space. 
We also consider a possible extension of the scenario by introducing other interactions with the SM particles. 

The rest of the paper is organized as follows. In \Sec{2}, we briefly explain the ALP inflation model. In \Sec{3}, we revisit the ALP miracle scenario by carefully studying the whole inflation parameter space to determine the viable parameter region.
In \Sec{matter}, we study the reheating process when the ALP has couplings with the SM fields other than photons. 
In \Sec{HD}, the production of ALP hot dark matter and dark radiation will be discussed focusing on 
thermalization condition of the ALP. The last section is devoted to discussion and conclusions.

\section{The ALP inflation}
\lac{2}
Let us first explain the inflation model where the ALP plays the role of inflaton. We assume that
the ALP enjoys a discrete shift symmetry, $\phi \to \phi + 2 \pi f$, where $f$ is the decay constant. 
Then, its potential is periodic: $V(\phi) = V(\phi+2\pi f)$.  Any periodic potential can be expanded as a sum of cosine terms.
If one of the cosine terms dominates the potential, it is the so-called
natural inflation~\cite{Freese:1990rb,Adams:1992bn}.  The model with multiple
cosine terms is called multi-natural inflation~\cite{Czerny:2014wza,Czerny:2014xja,Czerny:2014qqa}. 
As explained above, 
multiple cosine terms
are necessary to have successful inflation with a sub-Planckian decay constant.
In the simplest realization of the multi-natural inflation,
the potential consists of two cosine functions,
\begin{align}
\label{eq:DIV} 
V(\phi) = \Lambda^4\(\cos\(\frac{\phi}{f} + \theta \)- \frac{\kappa }{n^2}\cos\(\frac{n\f }{f }\)\)+{\rm 
const.},
\end{align}
where $n$ is an integer 
 greater than unity, $\theta$ and $\kappa$ respectively denote the relative phase and height of the two terms, and the last term is a constant required to realize the vanishingly small cosmological constant in the 
present vacuum. For $\kappa = 1$ and $\theta = 0$, the inflaton potential is reduced to the quartic hilltop inflation model.
In Ref.~\cite{Daido:2017wwb} we considered $\theta \ne 0$ while $\kappa$ was fixed to be unity.
In this paper,  to scan the full parameter space, we vary both $\theta$ and $\kappa$ and delineate the parameter space
where the successful slow-roll inflation takes place and they can give a better fit to the CMB data than the quartic hilltop 
inflation model~\cite{Takahashi:2013cxa,Czerny:2014wza,Daido:2017wwb}.

 If $n$ is an odd integer, the inflaton potential (\ref{eq:DIV}) exhibits a striking feature:
\begin{equation}
\laq{inv}
V(\phi + \pi f) = - V(\phi)+ {\rm const.},
\end{equation}
where the constant term is independent of $\phi$. Because of this feature, the curvatures at the potential maximum and minimum 
have the same absolute value but the opposite sign. During inflation, the potential must be sufficiently flat around the potential maximum 
for successful slow-roll inflation, and as a result, the inflaton remains light also at the potential minimum. In other words,
the inflaton can be long-lived because of the slow-roll condition in this case. In the following we focus on the case of $n$ being an odd integer, but our
analysis of the inflaton dynamics can be applied to the case of arbitrary $n$ without any changes.\footnote{One can slightly extend the model by replacing
the first term with $1/n'^2 \cos (n' \phi/f)$ with $n'$ being an odd integer smaller than $n$.}
We call $\phi$ the inflaton (ALP) when we discuss its dynamics during (after) inflation.

A flat-top potential with multiple cosine terms like (\ref{eq:DIV}) has several UV completions in extra dimensions; e.g., in the context of the 
extra-natural inflation with extra charged matters placed in the bulk~\cite{Croon:2014dma} and the elliptic inflation with the potential
given by an elliptic function obtained in the low-energy limit of some string-inspired set-up~\cite{Higaki:2015kta, Higaki:2016ydn}.
In both cases, it is possible to realize the potential similar to (\ref{eq:DIV}) with $n$ being an odd integer for a certain choice of the parameters.

During inflation the inflaton stays in the vicinity of the potential maximum,
 where one can expand the potential as
\begin{equation}
\laq{app}
V(\phi) = \frac{2(n^2-1)}{n^2} \Lambda^4 - \theta \frac{ \L^4}{f} {\phi}+  \frac{(\kappa-1)}{2}\frac{\L^4}{f^2} \phi^2
+ \frac{\h }{3!}\frac{\Lambda^4 }{f^3} {\phi^3} -  \frac{n^2-1 }{4!}\(\frac{\Lambda}{f}\)^4 \phi^4  + \cdots.
\end{equation}
Here we have included only terms up to the first order of $\h $ and $\kappa-1$, assuming they are much smaller 
than unity. In fact, the cubic term has a negligible effect on the inflaton dynamics for the parameters of our interest, 
and the potential can be well approximated by
\begin{equation}
\label{app}
V(\phi) \simeq V_0   - \h \frac{\L^4 }{ f} {\phi} +\frac{m^2 }{2}\f^2 - \lambda \phi^4,
\end{equation}
where we have defined 
 \begin{align}
 \label{eq:V0}
 V_0 \equiv 2\frac{n^2-1}{n^2} \Lambda^4,~~m^2\equiv {\(\kappa-1\)}\frac{\L^4 }{f^2},~~\lambda  \equiv \frac{n^2-1 }{4!}\(\frac{\Lambda}{f}\)^4.
\end{align}
Obviously, when $\kappa \rightarrow 1$ and $\h \rightarrow 0$, the potential (\ref{app}) has only a quartic term.

In the quartic hilltop inflation, the predicted scalar spectral index is given by $n_s \simeq 1-3/N_*$, where
$N_*$ is the e-folding number corresponding to the horizon exit of the CMB pivot scale, $k_*=0.05\,{\rm Mpc}^{-1}$.
Here and hereafter, the subscript $*$  implies that the variable is evaluated at the horizon exit of the pivot scale. 
Therefore the predicted scalar spectral index is considerably smaller than the observed value especially for $N_*< 50$.
In fact, the predicted value of $n_s$ can be increased to give a better fit to the CMB data by introducing 
a small but non-zero CP phase, $\theta$~\cite{Takahashi:2013cxa, Czerny:2014wza, Daido:2017wwb}.

The evolution of the inflaton can be expressed 
in terms of the e-folding number by solving
\begin{equation}
\laq{xecon}
N_* = \int^{\f_{\rm end}}_{\f_*}{\frac{H }{\dot{\f}} d \f} \simeq \int^{\f_{\rm end}}_{\f_*}{\frac{3H^2 }{-V'} d \f},
\end{equation}
where $H$ is the Hubble parameter, the overdot represents the derivative with respect to time, 
and we have used the slow-roll equation of motion, $3 H \dot{\phi} + V' \simeq 0$, in the second equality.
Here $\f_{\rm end}$ is the field value  at the end of inflation defined by $|\eta(\phi_{\rm end})|= 1$, 
and it is given by
\begin{align}
\label{phiend}
\phi_{\rm end} & \simeq \frac{2}{n} \frac{f^2}{M_{pl}} \ll \phi_{\rm min}.
\end{align}
Here $\phi_{\rm min} \sim \pi f$ is the inflaton field value at the potential minimum of \Eq{DIV},
and $M_{pl} \simeq 2.4 \times 10^{18}$\,GeV is the reduced Planck mass. We assume $\dot{\phi} > 0$ during inflation without loss of generality.
The e-folding number is fixed once the inflation scale and the thermal history after inflation are given.  As we will see, 
 the reheating takes place instantaneously and the energy for inflation immediately turns into radiation after inflation.
Assuming the instantaneous reheating, the e-folding number is given by
\begin{equation}
\laq{efold}
N_*\simeq 28 -\log{\(\frac{k_* }{0.05\, {\rm Mpc^{-1}}}\)}+\log{\(\frac{V_0^{1/4} }{10\,\TEV}\)}+\frac{1}{3} \log{ \(\frac{g_{*s}(T_0) }{3.95}\)}
-\frac{1}{12}\log{\(\frac{g_*(T_R) }{107.75 }\)},
\end{equation}
where
$g_*$ and $g_{*s}$ represent the effective number of relativistic species contributing to energy and entropy, respectively,
and $T_0$ and $T_R$ are the photon temperature at present and just after reheating, respectively. 
The reference values of $g_*$ and $g_{*s}$ incorporate 
 the contribution of thermalized ALPs (see discussion in  \Sec{HD}).

The power spectrum of the curvature perturbation is given by
\begin{equation}
\laq{pln}
P_{\mathcal{R}}(k_*) \simeq \(\frac{H_*^2}{2\pi \dot{\phi}_*}\)^2 \simeq 
\frac{V(\phi_*)^3}{12 \pi^2 V'(\phi_*)^2 M_{pl}^6},
\end{equation} 
and the CMB normalization condition reads
\begin{equation}
\label{pn}
P_{\mathcal{R}}(k_*) \simeq 2.2 \times 10^{-9},
\end{equation}
at $k_* = 0.05\, {\rm Mpc}^{-1}$\cite{Ade:2015lrj}. 
The scalar spectral index $n_s$ is given by
\begin{equation}
\laq{nsform}
n_s \simeq  1- 6 \varepsilon + 2\eta,
\end{equation}
where the slow-roll parameters are defined as,
\begin{align}
\varepsilon(\phi) &\equiv \frac{M_{pl}^2}{2} \left(\frac{V'}{V}\right)^2,\\
\eta({\f})  &\equiv M_{ pl}^2 \frac{V''}{V}.
\end{align}
In the small-field inflation, $\varepsilon(\phi_*)$ is much smaller than $|\eta(\phi_*)|$, and the spectral index is simplified to $n_s \simeq 1+ 2\eta(\phi_*)$.
In the numerical calculation we keep the terms up to the third order of the slow-roll expansion~\cite{Gong:2001he}.
The running of the spectral index is similarly given by
\begin{align}
\laq{run}
\frac{d n_s }{ d \log{k}}  & \simeq -24 \varepsilon^2 + 16 \varepsilon \eta  - 2 \xi \simeq -2 \xi,
\end{align}
where we have defined 
\begin{equation}
\xi(\phi) \equiv  M_{pl}^4 \frac{V' V'''}{V^2}.
\end{equation}
In the second equality of \eq{run} we have dropped the terms containing $\varepsilon$. The running of the spectral index is
usually small except for the modulated potential~\cite{Kobayashi:2010pz}, but it can be sizable since $N_*$ is much smaller
than the conventionally used values $50$ or $60$.

Now let us turn to constraints from the CMB observation.
We adopt the following constraints on $n_s$ and its running obtained by Planck TT+lowP+lensing~\cite{Ade:2015lrj}, 
\begin{align}
\label{eq:case1}
n_s(k_*)& =0.968\pm 0.006,  \\
\laq{crun}
\frac{d{n_s}}{d\log{k}}(k_*) & = -0.003 \pm 0.007, 
\end{align}
where  the running of the running of the spectral index, $d^2 n_s/d (\log{k})^2$,  is set to be zero. If the running of the running is
allowed to float, the above constraints will be slightly modified. However, according to Ref.~\cite{Cabass:2016ldu}, 
the preferred range of the running of running is biased to positive values, and its $2 \sigma$ lower bound reads 
$d^2 n_s/d (\log{k})^2 > - 0.001$. On the other hand, in our model, the running of the running takes only negative values for
the parameters of our interest, and as long as it satisfies the above $2 \sigma$ lower bound, its effect on the
running and $n_s$ is negligibly small. Therefore we can justifiably apply the constraints (\ref{eq:case1}) and \eq{crun} to our model, if
we limit ourselves to the region with $d^2 n_s/d (\log{k})^2 > - 0.001$.

\begin{figure}[!t]
  \begin{center}
   \includegraphics[width=105mm]{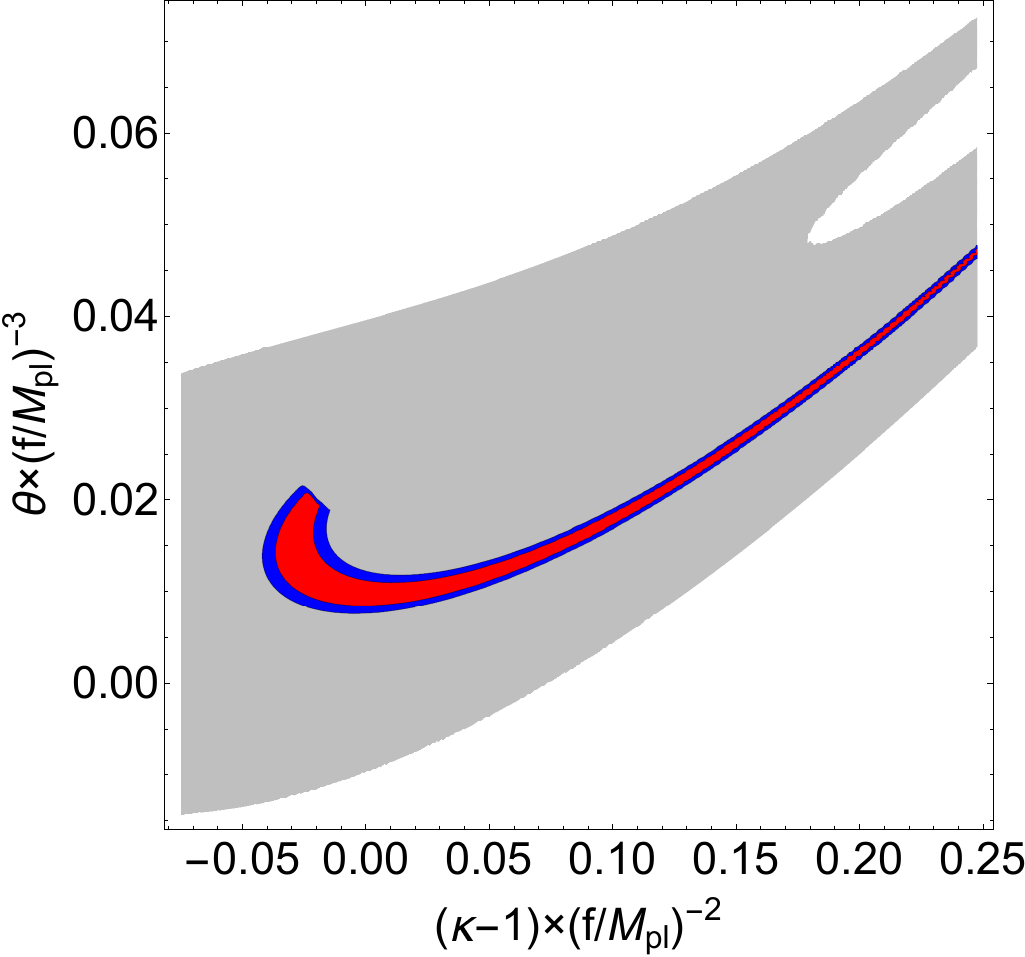}
\end{center}
\caption{
In the red and blue shaded regions the CMB constraints (\ref{eq:case1}) are satisfied at $1\sigma$ and $2\sigma$, respectively. 
We have set $n=3$ and  $f=10^7\GEV$, and normalized $\theta$ and $\kappa - 1$ by their typical values for convenience. 
In the gray region $\ab{\eta(\phi_*)}<0.1$ is satisfied. }
\label{fig:ns}
\end{figure}

{
We have numerically solved the inflaton dynamics using the potential \eq{app},
varying both $\theta$ and $\kappa-1$ around $0$.
Here and in what follows $n=3$ and $f = 10^7\,$GeV are chosen as reference values,
unless otherwise stated.
As we shall see later,  while larger values of $f$ are consistent with inflation and the CMB data, 
$f$ should be close to $10^7$\,GeV for successful reheating and explanation of dark matter.
}
In Fig.~\ref{fig:ns}, we show the viable parameter region in the $(\kappa,\h)$ plane, 
where the red and blue regions correspond to the $1\s$ and $2\s$ of (\ref{eq:case1}) and \eq{crun}. For simplicity we have not taken into account of the correlation between $n_s$ and its running.
 The left side of the viable region is bounded from above around $\theta \sim 0.02 (f/M_{pl})^3$, because
we have imposed a condition $d^2 n_s/d (\log{k})^2 > -0.001$. {We have confirmed that the running of the spectral index
tends to be larger than { the conventional case with $N_* = 50$ or $60$ } because of the small e-folding number.}
In the gray shaded region,  $|\eta(\phi_*)| < 0.1$ is satisfied,  where the curvature of the potential is so small that slow-roll inflation 
takes place. { We note that the viable region lies in a tiny range of $\kappa$ for a fixed $\theta$, which requires 
some amount of fine-tuning of the parameters for successful slow-roll inflation. In the scope of this paper, we do not pursue the
origin of such fine-tuning, but we note that a similar flat-top potential arises from extra-dimensional 
set-up~\cite{Croon:2014dma,Higaki:2015kta, Higaki:2016ydn}.}

One can see that the viable region continues to larger values of $\kappa$ and $\h$. In the limit of large $\kappa-1$ and $\theta$,
the inflation takes place around a saddle point, not the maximum. We do not consider this possibility further
in this paper because successful reheating becomes so inefficient that the relic inflaton condensate exceeds
the observed dark matter abundance. Note also that, in the limit of the saddle-point inflation, the inflaton mass during inflation is
not directly related to that at the potential minimum, which breaks the link between low-energy observables 
and the inflaton parameters.

\begin{figure}[t!]
\begin{center}  
   \includegraphics[width=105mm]{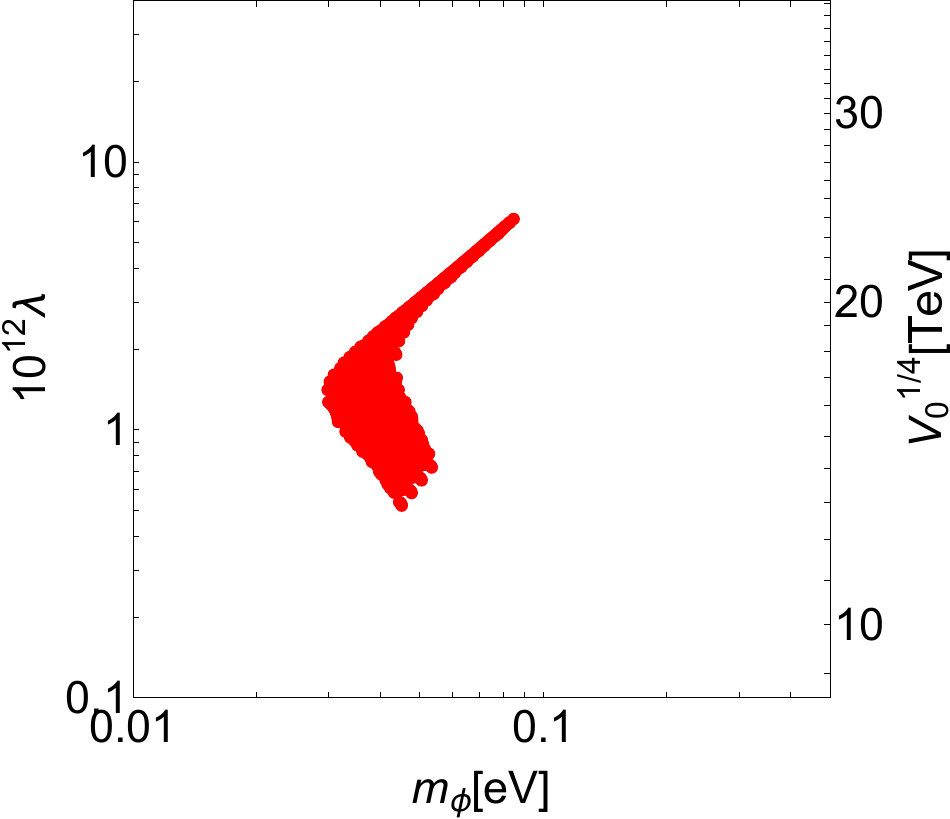}
\end{center}
\caption{The range of $m_\f$ and $\l$ with $f=10^7\GEV$. The points are consistent with the CMB observation at more than $95\%$ CL.) }
\label{fig:mlambda}
\end{figure}

Now let us evaluate the inflaton parameters at the potential minimum using the relation of \Eq{inv}. 
To avoid the confusion let us call $\phi$ the ALP when we discuss its properties around the minimum.
The potential can be similarly expanded around the minimum as
\begin{equation}
\laq{pbm}
V(\phi + \pi f) \simeq   \h \frac{\L^4 }{ f} {\phi} -\frac{m^2 }{2}\f^2+\lambda \phi^4+ \cdots.
\end{equation}
Let us denote the field values of $\phi$ at the potential maximum and minimum as
$\f_{\rm max}$ and $\phi_{\rm min}$, respectively. They satisfy $\f_{\rm min}=\f_{\rm max}+\pi f$ because of \Eq{inv}.
We can then estimate the ALP mass, $m_\phi$, as well as the self-coupling, $\lambda$, at $\phi_{\rm min}$ in the viable region shown in Fig.~\ref{fig:ns}. The result is shown in Fig.~\ref{fig:mlambda}.
Here, the upper bounds on $\l$ and $m_\f$ are set by the constraints on the running of the 
running of the spectral index, while the lower bound of $\l$ is determined by the largest of $\kappa$ in Fig.~\ref{fig:ns}.
As $\kappa$ becomes larger, $\lambda$ becomes smaller, which makes it more difficult to reheat the Universe because
the inflation scale becomes lower. Therefore we focus on the parameters shown in Fig.~\ref{fig:mlambda} in the following.
One can see that the ALP mass and its self-coupling are in the following ranges:
\begin{align}
\label{mass}
0.03\,\EV &\lesssim m_\f \lesssim 0.1\,\EV,\\
5 \times 10^{-13} &\lesssim \l  \lesssim 7 \times 10^{-12},
\end{align}
for $f= 10^7\GEV$. 
Note that the ALP mass $m_\phi$ is of order the Hubble parameter during inflation. This is because
of the observational constraint on the spectral index. To see this, let us rewrite $m_\phi$ as 
\begin{equation}
\laq{sim}
m_\f^2 \equiv {V'' }(\f_{\rm min}) =  \left|V''(\f_{\rm max})\right|
\sim \left|V''(\f_*)\right| \simeq \frac{3}{2} \left| n_s-1 \right| H^2_*,
\end{equation}
The equality in the middle reflects the fact that the inflation takes place around the potential maximum.
So, substituting the observed value $1-n_s \simeq 0.03$, one obtains  $m_\phi \sim H_*$. 

In the quartic hilltop inflation, it is $\lambda$ that is fixed by the CMB normalization (\ref{pn}) 
and it only depends logarithmically on the inflation scale through the e-folding number, $\l \propto N_*^{-3}$.
Therefore, $\l \sim \O(10^{-12})$ holds for a broad range of $f$.
Since $H_*^2 \propto  \l f^4$,  the ALP mass and decay constant scale as $m_{\f}\propto f^{2}$.

We stress that the relation \eq{sim} holds for a broader class of the ALP inflation (with e.g. more cosine terms) 
satisfying \eq{inv}, as long as slow-roll inflation takes place in the vicinity of the potential maximum.
This is partly because $V'''(\f_*)$ is tightly constrained by the observation as it contributes to the running as well as the running of 
the running of the spectral index. Our argument on the reheating and the relic ALP abundance in the rest of this paper relies
on the two relations \eq{sim} and $\l \sim \O(10^{-12})$, and so, we expect that our results in the following sections are not
significantly modified for a broader class of the ALP inflation model.

\section{The ALP miracle}
\lac{3}
The inflaton (ALP) is light at both the potential maximum and minimum in a class of the ALP inflation,
in which case the ALP becomes long-lived and therefore can be dark matter. As we shall see shortly, 
such axionic unification of inflaton and dark matter is not only economical but also has interesting 
implications for future axion search experiments and the small-scale structure problem. 

\subsection{Coupling to photons and dissipation rates}
After inflation, the ALP starts to oscillate about the potential minimum. For successful reheating, one needs to 
couple the ALP to the SM fields. Here we introduce the ALP couplings to weak gauge bosons,
which is reduced to a coupling to photons in the low energy. Other interactions such as couplings to fermions will be
studied in the next section.   
At first, it seems that the reheating is hampered by the suppressed decay rate in the vacuum, but as we shall see, 
the effective mass of the ALP condensate soon after inflation is much larger than the ALP mass in the vacuum. 
After thermal plasma is generated by the perturbative decay, the reheating mainly proceeds through thermal dissipation process.

Let us consider the ALP coupling to photons,
\begin{align}
\laq{int}
{\cal L} & = c_\g \frac{\a}{4 \pi} \frac{\phi}{f} F_{\mu \nu} \tilde F^{\mu \nu} 
\equiv \frac{1}{4} g_{\phi\g\g} \phi F_{\mu \nu} \tilde F^{\mu \nu}
\end{align}
 where
$c_\g$ is a model-dependent constant and $\a$ is the fine structure constant.
The ALP condensate decays into a pair of photons at the decay rate given by\footnote{Strictly speaking, 
one should use the decay rate averaged over time during the oscillation. 
However, we have confirmed that this does not change our results, as the dominant process for the 
reheating is the dissipation process. }
\begin{equation}
\laq{dec}
\Gamma_{\rm dec,\g} =\frac{ c_\g^2\alpha^2 }{64 \pi^3} \frac{m_{\rm eff}^3}{f^2}.
\end{equation} 
where we have introduced the effective mass of the ALP as 
\begin{equation}
\laq{eff}
m_{\rm eff}^2 \equiv 12\l \f_{\rm amp}^2=12 \sqrt{\l \r_\f},
\end{equation}
given in terms of the oscillation amplitude measured from the potential minimum, $\f_{\rm amp}$, 
and the oscillation energy, $\r_\f$, of the ALP condensate.  
Note that the potential about the minimum is dominated by the quartic term in \Eq{pbm},
and the mass term is negligible until the oscillation amplitude becomes sufficiently small.
The effective mass decreases with time as the oscillation 
energy is red-shifted. In the present Universe, the oscillation amplitude of the relic ALP condensate is so small that
the potential is well approximated by the quadratic term. We emphasize here that the decay rate of the ALP just after inflation 
is significantly enhanced compared to the one in the vacuum.

\begin{figure}[t!]
\begin{center}  
   \includegraphics[width=110mm]{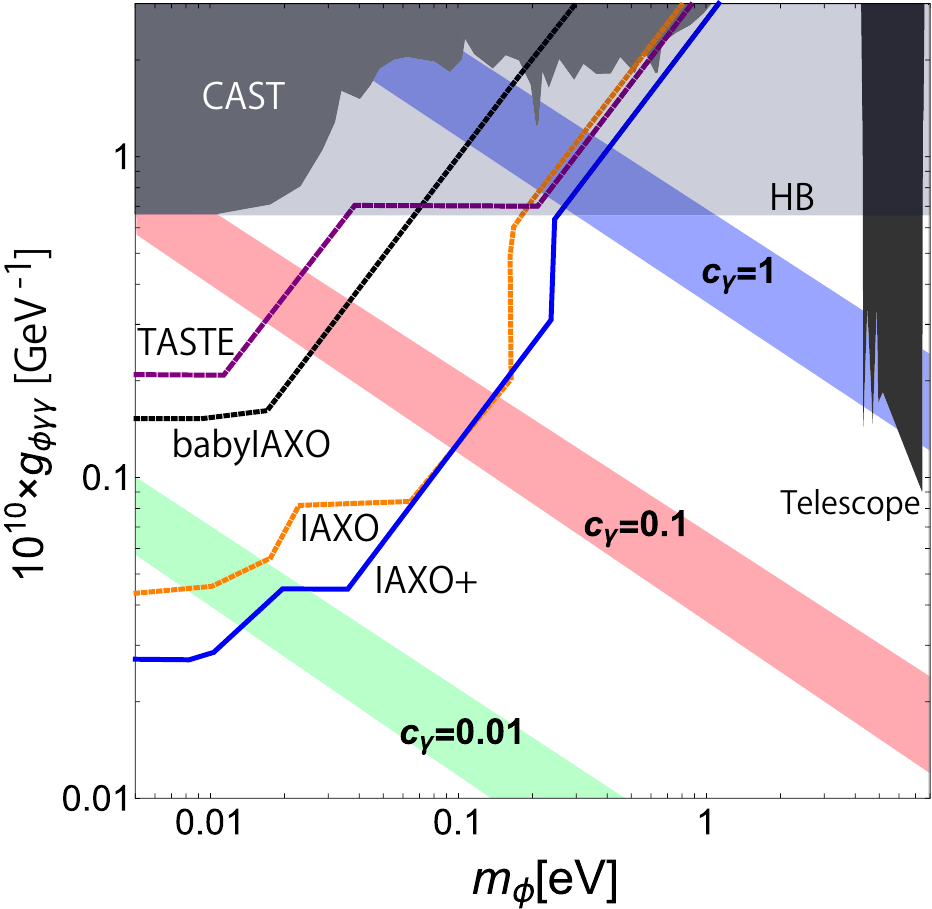}
   \end{center}
\caption{The preferred region of $g_{\f\g\g}$ and $m_\f$ for  $c_\g = 1$ (blue), $0.1$ (red), and $0.01$ (green).
The gray, light gray, and black regions are excluded by CAST, the cooling arguments of HB stars, and 
the optical telescopes, respectively~\cite{Anastassopoulos:2017ftl, Ayala:2014pea, Grin:2006aw}. 
The last constraint assumed the ALP dark matter.
The purple, black, orange, and blue lines show the projected sensitivity reach of TASTE, baby IAXO, IAXO, and IAXO+, respectively\cite{Irastorza:2011gs, Armengaud:2014gea, IAXO, Anastassopoulos:2017kag, Giannotti:2017hny}. 
 }
\label{fig:phig}
\end{figure}

  In Fig.~\ref{fig:phig} we show the relation between $m_\phi$ and $g_{\phi \gamma \gamma}$
for $c_\gamma = 0.01, 0.1,$ and $1$ as diagonal bands from bottom to top. Each band corresponds to the 
viable parameters consistent with the CMB observations for different values of $f$. (The case of $f = 10^7$\,GeV is shown in Fig.~\ref{fig:ns}).
We also show the regions constrained by the CAST experiment~\cite{Anastassopoulos:2017ftl}, the cooling argument of the horizontal branch (HB) stars~\cite{Ayala:2014pea}
and the optical telescopes (assuming the ALP dark matter)~\cite{Grin:2006aw}, and the projected sensitivity reach of the future experiments.
One can see that a large fraction of the viable parameter space consistent with the CMB observation will be covered by the future experiments.

Soon after a small fraction of the ALP condensate decays into photons, the produced photons quickly form thermal 
plasma, and its back reactions become relevant.
One of the back reactions is the thermal blocking effect. Photons in the thermal plasma acquire
a thermal mass of order $eT$, and once it exceeds half the ALP effective mass, the perturbative decay is kinematically 
forbidden. Then, evaporation process of the ALP condensate with the ambient plasma such as $\f+\g \rightarrow e^{-}+e^{+}$
 becomes significant~\cite{Yokoyama:2005dv,Anisimov:2008dz,Drewes:2010pf,Mukaida:2012qn,Mukaida:2012bz}. 
The thermal dissipation of the QCD axion  was studied in Ref.~\cite{Moroi:2014mqa}, where it was pointed out that 
the dissipation rate is accompanied by a suppression
factor, $p^2/g_s^4 T^2$, when the typical momentum of the axion, $p$, is smaller than $g_s^2 T$, where $g_s$ is the
strong gauge coupling. In the case of spatially homogeneous condensate, the momentum should be replaced with the
(effective) mass. In our case, even though the ALP condensate is (almost) spatially homogeneous just after inflation, 
inhomogeneities quickly grow due to the tachyonic preheating~\cite{Felder:2000hj,Felder:2001kt}.
The initial typical peak momentum is about the effective mass or less~\cite{Brax:2010ai}.
Afterwards, the finite momentum modes continue to scatter via the quartic coupling, 
and the spectrum gets broader leading to self-similar evolution~\cite{Micha:2004bv}. As a result,
the typical momentum of the ALP condensate can be larger than the effective mass by a factor of several.\footnote{
See Ref.~\cite{Lozanov:2017hjm} for the latest detailed analysis on the scalar resonance. We thank M. Amin for useful communication on this issue.
}
Applying the result of Ref.~\cite{Moroi:2014mqa} to the ALPs,
we obtain the following dissipation rate, 
\begin{equation}
\laq{disIR}
\Gamma_{{\rm dis,\g}} = C  \frac{c_\g^ 2 \alpha^2T^3}{8 \pi^2f^2} \frac{m_{\rm eff}^2}{e^4T^2},
\end{equation}
where $C$ is a numerical constant of ${\cal O}(10)$ which represents the uncertainties of the
order-of-magnitude estimation of the dissipation rate as well as the effect of tachyonic preheating and the 
scalar resonance~\cite{Felder:2000hj, Felder:2001kt, Micha:2004bv, Lozanov:2017hjm}.

At temperatures higher than the electroweak scale, $T>T_{\rm EW}$, one should
consider the ALP coupling to SU(2)$_L$ and U(1)$_Y$ gauge bosons:
\begin{align}
\laq{aWWBB}
{\cal L} & = c_2 \frac{\a_2}{8 \pi} \frac{\phi}{f}W^a_{\mu \nu} {\tilde W}_a^{\mu \nu}
+ c_Y \frac{\a_Y}{4 \pi} \frac{\phi}{f} B_{\mu \nu} \tilde B^{\mu \nu},
\end{align}
where $(\alpha_2, W^a_{\mu \nu})$ and $(\alpha_Y, B_{\mu \nu})$ are the fine-structure constants and gauge field strengths of SU(2)$_L$ and U(1)$_Y$, respectively, and $c_2$ and $c_Y$ are anomaly coefficients.
For instance, if there are extra fermions $\p_i$ in the fundamental representation of
the SU(2)$_L$, $c_2$ is given by
\begin{equation}
c_2 = \sum_i q_i,
\end{equation}
where $q_i$ is the charge of $\psi_i$ under the shift symmetry. 
Similarly, $c_Y$ is given by
\begin{equation}
c_Y = \sum_i q_i Y_i^2,
\end{equation}
where $Y_i$ is the hypercharge of the $i$-th  chiral fermion.

In the low energy, the above interactions are reduced to the coupling to photons \Eq{int}, and
the coefficients satisfy
 \begin{equation}
 \label{eq:cgcwcy}
 c_\gamma = \frac{c_2}{2} + c_Y.
 \end{equation}
The dissipation rate due to the couplings to the weak gauge bosons is similarly given by~\cite{Daido:2017wwb, Moroi:2014mqa}
\begin{equation}
\Gamma_{{\rm dis,EW}} = C' \frac{c_2^2 \alpha_2^2T^3}{32\pi^2f^2} \frac{m_{\rm eff}^2}{g_2^4T^2}
+C'' \frac{c_Y^2 \alpha_Y^2T^3}{8 \pi^2f^2} \frac{m_{\rm eff}^2}{g_Y^4T^2}, \laq{disEW}
\end{equation}
where $C'$ and $C''$ are constants of ${\cal O}(10)$.

Taking account of the above thermal effects, we  obtain the following  Boltzmann
equations, 
\begin{align}
\left\{\begin{array}{ll}
	\displaystyle{\dot{\rho}_\phi+4H\rho_\phi=-\Gamma_{\rm tot}\rho_\phi} \\
	&\\
	\displaystyle{\dot{\rho}_r+4H\rho_r=\Gamma_{\rm tot}\rho_\phi}\label{evolution}
	\end{array}
	\right.,
\end{align}
where $\rho_r$ denotes the energy density of the radiation. Note that the ALP energy density decreases like radiation
since it oscillates in a quartic potential. 
The total dissipation rate is given by 
\begin{align}
\G_{\rm tot}=\left\{\begin{array}{ll}{\displaystyle{ \G_{\rm dec,\g}+\G_{\rm dis,\g}}~ (T<T_{\rm EW})}\\ 
&\\ 
\displaystyle{
\G_{\rm dis,EW}~(T>T_{\rm EW})} 
\end{array} \right..
\end{align}
In numerical calculations we also take account of the perturbative ALP decay to the weak bosons if kinetically allowed,
but this does not change our results since the dominant process is due to the dissipation  at $T>T_{\rm EW}$.
We adopt the initial condition (at $t=t_i$) as
\begin{equation}
{\r_\f}|_{t=t_i}= V_0\simeq \frac{48 }{n^2}\l f^4,~~ \rho_r|_{t=t_i}= 0,
\end{equation}
where we have set the initial oscillation energy equal to the total energy just after the inflation.\footnote{Although the potential \eq{DIV} can not be approximated by the potential \eq{pbm} soon after inflation, the tachyonic preheating takes place to reduce the amplitude during the first a few oscillations to be within the region $\f_{\rm amp}< f \pi/2$ where the potential is well approximated by \eq{pbm}. 
} 

The dissipation becomes ineffective at a certain point since the dissipation rate decreases faster than the Hubble parameter.
 The fraction of the relic ALP energy density to the total energy density becomes constant afterward, and 
we denote it by
\begin{equation}
\xi \equiv \left.\frac{\rho_\phi}{\rho_\phi + \rho_r}\right|_{\rm after~dissipation}.
\end{equation} 
One can show from Eq.~(\ref{evolution}) that the ratio asymptotes to a constant when $\G_{\rm tot}$
becomes smaller than $H$. 

For the parameters of our interest, the ALP condensate behaves like radiation during the big
bang nucleosynthesis (BBN) epoch, and $\xi$ is related to the extra effective neutrino species, $\Delta N_{\rm eff}$, as
\begin{align}
\frac{\xi}{1-\xi} = \frac{7}{4} \frac{g_*(T_R)^\frac{1}{3}}{g_*(T_{\rm BBN})^\frac{4}{3}} \Delta N_{\rm eff},
\end{align}
where $T_R$ is the reheating temperature and $T_{\rm BBN} = {\cal O}(1)$\,MeV.
 For instance, in order to satisfy $\Delta N_{\rm eff} < 1$ during the BBN era, $\xi$ must be smaller than about $0.26$. 
As we shall see shortly, an even tighter constraint on $\xi$ is required for explaining the dark matter abundance by the relic ALP condensate.

\subsection{The ALP miracle}

Before presenting the numerical results, let us make a rough estimate of the ALP relic abundance and
discuss various constraints on the ALPs. We will see that the ALP mass and coupling to photons suggested by
the axionic unification of inflaton and dark matter happen to be close to the current experimental and observational 
constraints. Interestingly, there are also some anomalies which can be interpreted as a hint for ALPs 
around the current bounds. We call this non-trivial coincidence {\it the ALP miracle}.

{
The ALP relic abundance is fixed when the evaporation rate becomes comparable to and
 smaller than the Hubble parameter,
\begin{equation}
\laq{fro}
H \gtrsim  \G_{\rm dis,EW}.
\end{equation}
The dissipation rate should not be much larger than the Hubble parameter however, since
otherwise the ALP condensate evaporates almost completely.
Then, let us consider a case where the above equality holds just after inflation and the dissipation
stops within a few Hubble time.
 Assuming the dissipation rates into SU(2)$_L$ and U(1)$_Y$ gauge bosons are of the same order, one can rewrite 
 this condition as
\begin{equation}
\laq{gaugediss}
\x\sim  0.004  \(\frac{30 }{C''}\)^2  \(\frac{g_*(T_R) }{107.75 }\)^{1/2} \(\frac{10^{-12}}{\l}\)^{3/2}  \(\frac{f/c_Y^2}{10^7\GEV}\)^2,
\end{equation}
where we have used the Friedmann equation $H = \sqrt{\rho_\phi + \rho_r}/\sqrt{3} M_{pl}$ and \Eq{eff},
and assumed that the dissipation completes soon after inflation ends.
}
On the other hand, the bounds on $g_{\phi \gamma \gamma}$ from CAST and HB  stars\cite{Anastassopoulos:2017ftl,Ayala:2014pea} 
are given by 
\begin{equation} 
\laq{cast} g_{\f\g\g} <
0.66 \times 10^{-10}\GEV^{-1} 
 \end{equation} 
or equivalently,
$$
f/c_\g \gtrsim 3.5 \times 10^{7}\GEV.
$$
Here the above CAST bound is valid for $m_\phi < 0.02$\,eV and it is weaker for a heavier ALP mass.
This implies that $\xi$ is of ${\cal O}(0.01)$ for $f\sim 10^7$\,GeV, $c_\g \sim 0.3$,
$c_Y \sim 1$ and $C'' \sim 30$, while marginally satisfying the current bound on $g_{\phi \gamma \gamma}$.

After the dissipation becomes inefficient, the oscillation amplitude $\phi_{\rm amp}$ continues to decrease inversely proportional 
to the scale factor due to the cosmic expansion. Therefore, the energy density of the relic ALP condensate behaves like radiation until the quadratic 
term becomes relevant. The ALP condensate behaves like non-relativistic matter afterward. 
The oscillation amplitude when the quadratic potential becomes equal to the quartic one is given by  $\phi_{\rm f} = m_\phi/\sqrt{2 \lambda}$.
The red-shift parameter at the transition is given by
\begin{equation} 
\laq{zc}
1+z_{\rm f}= \(\frac{1}{\Omega_{\f}} \frac{\rho_{\rm f}  }{ \r_{c}}\)^{\frac{1}{3}} \sim 7\times 10^{5} \(\frac{0.12 }{\Omega_{\f}h^2}\)^{\frac{1}{3}} \(\frac{10^{-12} }{\l}\)^{\frac{1}{3}} \(\frac{m_\f }{0.05\EV}\)^{\frac{4}{3}},
\end{equation}
where $\Omega_{\f}$ is the density parameter of the relic ALP condensate,
$h$ is the reduced Hubble parameter,  $\r_{c}$ is the current critical energy density, $\rho_{\rm f} \simeq m_\f^4 /2 \l$ is 
the ALP energy density at the transition\footnote{
Precisely speaking, the ALP condensate has the gradient energy due to the tachyonic preheating and the scalar resonance.
This could slightly delay the transition but does not significantly change our discussion. }.
If the ALP plays the role of dark matter,  the matter power spectrum at small scales is suppressed 
since the transition takes place at a relatively late time.
The transition of such  late-forming dark matter is constrained by the SDSS (Ly-$\a$) data as \cite{Sarkar:2014bca},
\begin{equation}
\laq{lya}
1+z_{\rm f}\; \gtrsim\;  1 \times 10^5~(9 \times 10^5)~~~(99.7\% {\rm CL}).
\end{equation}
If we adopt the SDSS bound, this implies a lower bound on the ALP mass, 
\begin{equation}
m_\f\gtrsim 0.01\EV.
\end{equation}

The relic ALP abundance can be estimated as follows. The present ratio of the ALP energy density to entropy  is
given by
\begin{align}
\left.\frac{\rho_\phi}{s}\right|_0 = \left.\frac{\rho_\phi }{s}\right|_{\rm after~dissipation} \times \frac{\phi_{\rm d}}{\phi_{\rm f}}
\end{align}
where $\phi_{\rm d} \simeq (4 \xi V_0/\lambda)^{1/4}$ represents the oscillation amplitude 
when the dissipation is decoupled. {The last factor represents the fact that $\rho_\phi/s$ decreases
inversely proportional to the scale factor before the transition.}
Assuming the instantaneous reheating, we obtain
\begin{align}
\label{eq:mbelow}
\Omega_{\f}h^2 \simeq  0.07  
 \(\frac{g_*(T_R) }{107.75 }\)^{-\frac{1}{4}} 
\left(\frac{\xi}{0.01}\right)^{\frac{3}{4}} \(\frac{10^{-12} }{\l}\)^{1/4} \left( \frac{m_\f }{0.05\EV} \right) ,
\end{align}
where we have used the approximation $\xi \ll 1$. 

To sum up, the successful inflation and explaining dark matter by a single ALP led us to a particular parameter region, 
 $m_\phi = {\cal O}(0.01)$\,eV and $\lambda \sim 10^{-12}$ 
for $f \sim 10^7$\,GeV (see Eq.~(\ref{mass})). These ALP mass and decay constant are the reference values
for our scenario. Introducing the ALP coupling to photons, we have shown that the reheating occurs 
mainly through dissipation effect. To be explicit, the fraction of the ALP energy density can be reduced
to $\xi = {\cal O}(0.01)$ for the reference values. Interestingly the relic ALP abundance turns out to be
close to the observed dark matter abundance for the same parameters. We have also seen that the suggested ALP 
mass and coupling to photons marginally satisfy the CAST and HB star bounds as well as the small-scale structure constraint,
and therefore the suggested parameters can be probed by the future experiments and observations. 
We call such coincidences the ALP miracle. 

\vspace{5mm}

To confirm the ALP miracle, we have numerically solved the Boltzmann equations (\ref{evolution}) for the viable parameters consistent
with the CMB observation, varying unknown numerical coefficients by a factor of $\O(1)$. To be concrete, we first 
collect about $100$ sets of $(\kappa, \theta)$ in the viable region consistent with the CMB observation at 2 $\s$ level for each value of $f$. The ranges of $\kappa$ and $\h$ are taken to be $-0.04 < (\kappa-1)\(f/M_{\rm pl}\)^{-2}< 0.25$ and 
$0 <\h \(f/M_{\rm pl}\)^{-3}< 0.08 $, and we vary $f$ in the range of $10^6\GEV < f < 5\times10^7 \GEV$ with
an interval of $5\times 10^5\GEV$.
For each point we randomly generate 50 sets of $(c_2, c_Y)$ in the range of $0<c_2<5$ and $|c_Y|<5$.
Thus, each point is characterized by five parameters $(f, m_\f, V_0, c_2, c_Y)$. For a given set of the input parameters, 
we numerically solve the Boltzmann equation (\ref{evolution}) and estimate the relic ALP abundance $\Omega_{\f}h^2$ 
from \Eq{mbelow} by setting the prefactors $C=C'=C''$ equal to $10$ \AND $30$. 

\begin{figure}[t!]
\begin{center}  
   \includegraphics[width=75mm]{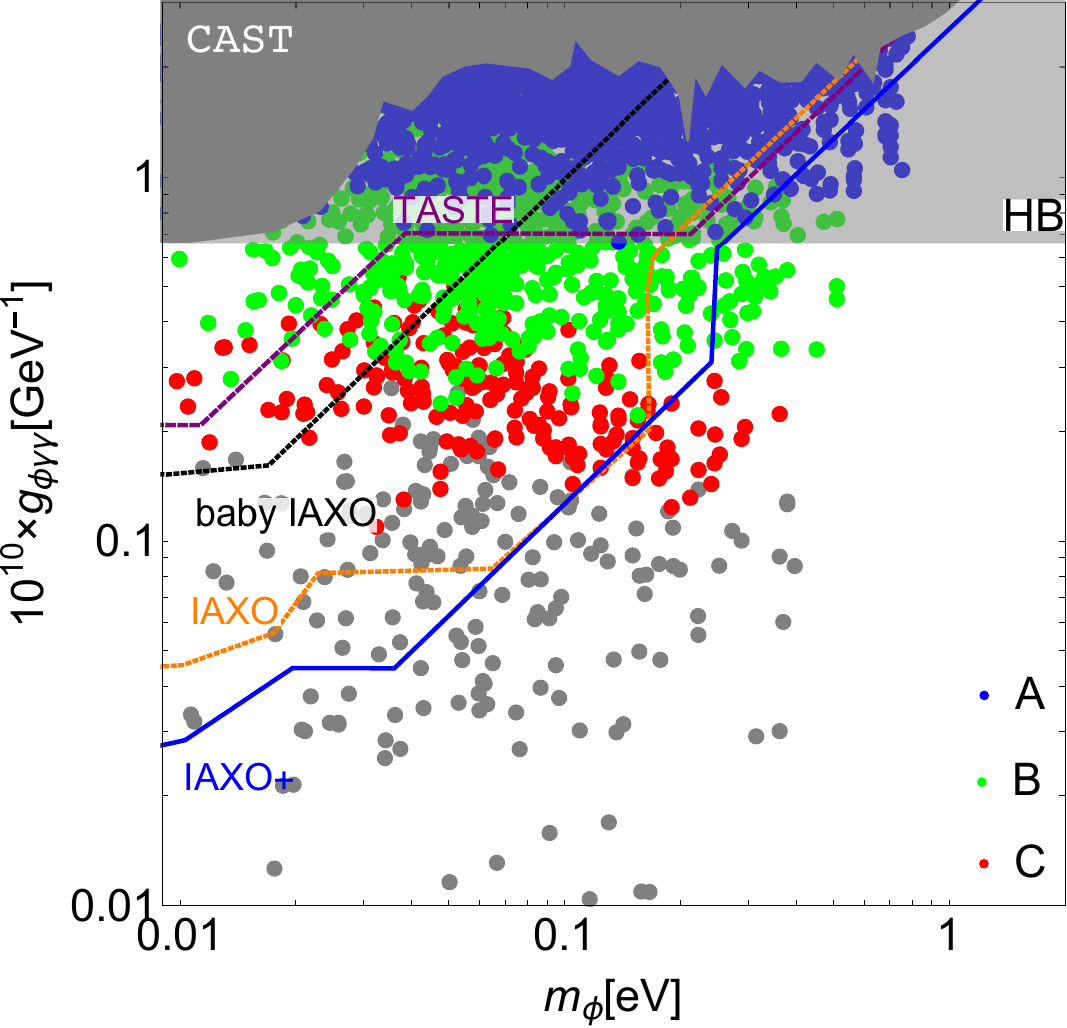}
   \includegraphics[width=75mm]{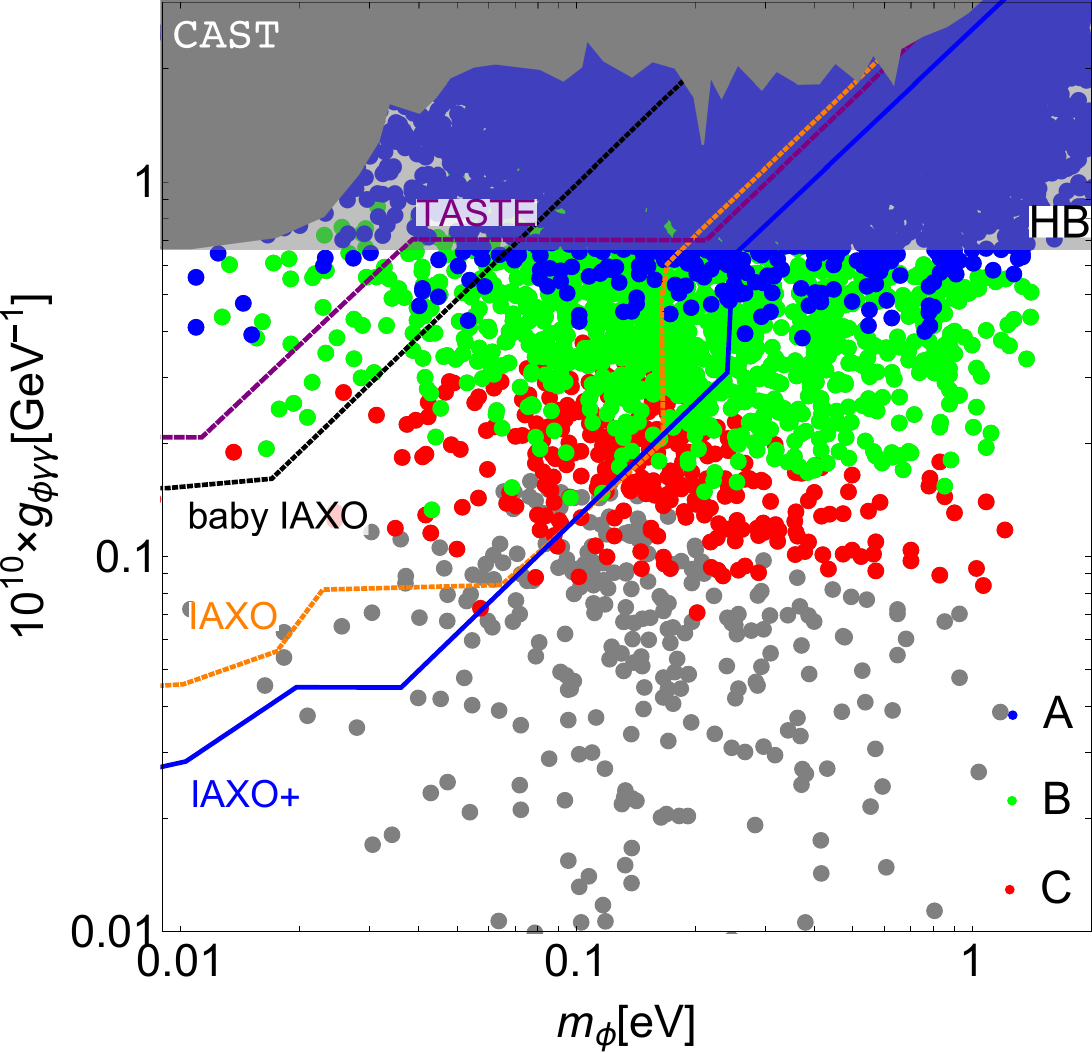}
\end{center}
\caption{Scatter plot of the viable parameters consistent with the CMB observation and the dark matter abundance for $C=10$ (left panel) and $C=30$ (right panel).
The colors of the points correspond to the ratio of the ALP couplings to photons and hypercharge gauge bosons (see the text).
The constraints, as well as the projected sensitivity reaches, are same as in Fig.~\ref{fig:phig}. 
 }
\label{fig:gauge}
\end{figure}

We show those points satisfying $0.08<\Omega_{\f} h^2<0.16$ in Fig.~\ref{fig:gauge},
where we divide the parameter sets into three groups according to the ratio,
$
\d= \ab{c_\g /c_Y}.
$
The groups A, B and C represent those points with  $0.3<\d$, $0.1 < \d < 0.3$ and $0.05 < \d < 0.1$, respectively.
The gray points with $\d <0.05$ require certain choices of the matter contents and charge assignment but 
do not necessarily imply fine-tuning of the parameters.
For instance, $\d=0$ is realized for $c_2=1$ and $c_Y=-1/2$.\footnote{In this case, the ALP should be heavier than $\sim 1 \EV$ so that the effective mass can be larger than twice of the weak boson mass to decay into 
SM particles instantaneously. 
Since it does not couple to photons at low energy,
it can not be probed in IAXO or photon-photon collider, while it still can be tested in the light of the CMB and BAO observation 
since the ALP becomes hot dark matter as in \Sec{HD}. } 
The gray (light gray) region is excluded by the constraints from CAST (HB stars). 
One can see that there are viable parameters satisfying the current bounds (mostly groups B and C), and that
a significant fraction of them can be probed by the future solar axion helioscope experiments.

\begin{figure}[t!]
\begin{center}  
   \includegraphics[width=75mm]{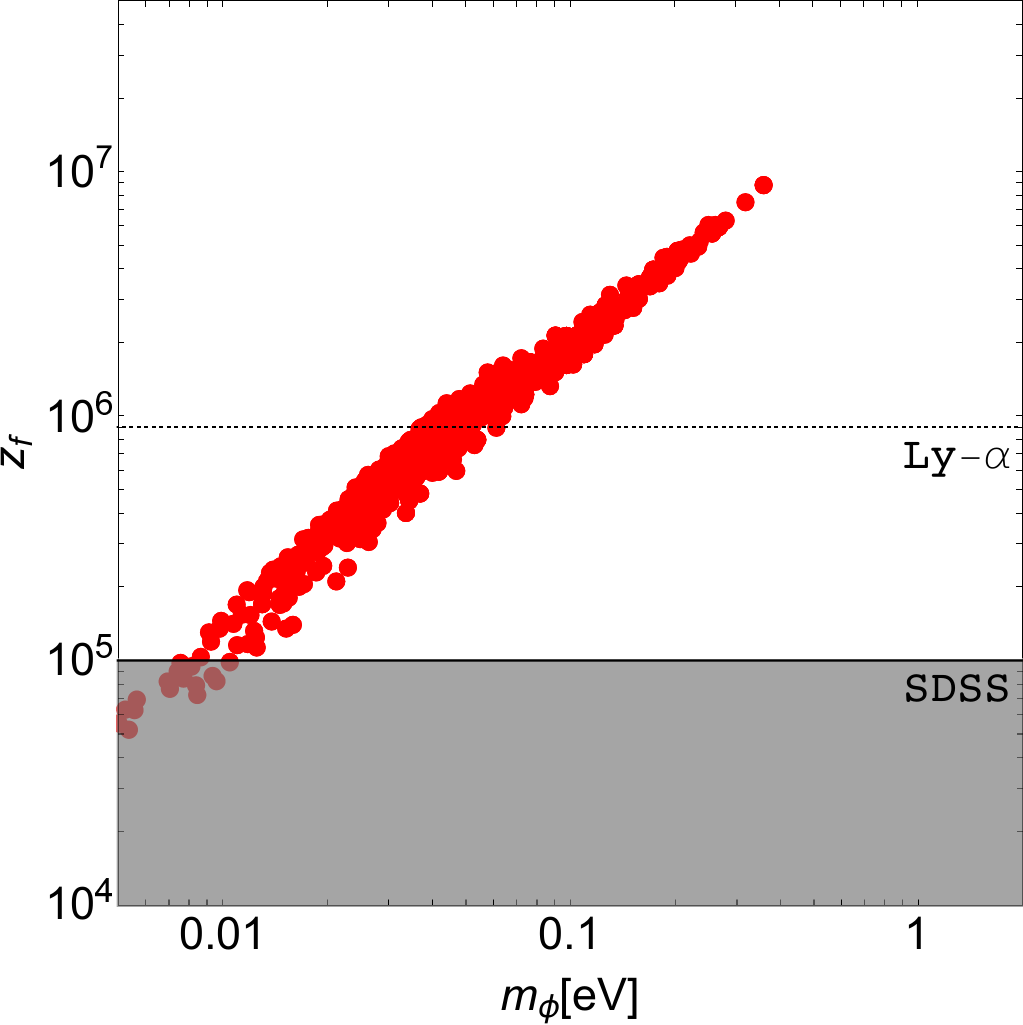}
   \includegraphics[width=75mm]{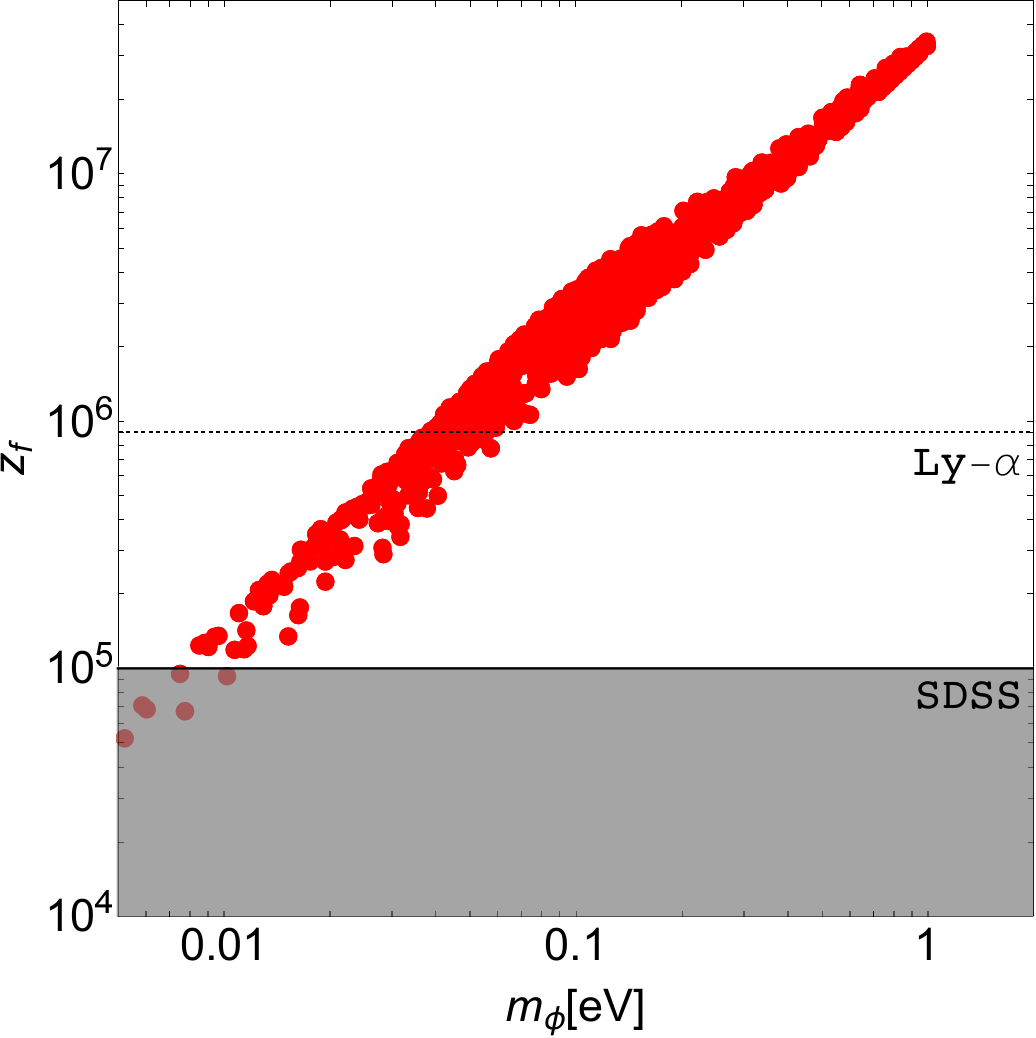}
\end{center}
\caption{Scatter plot of $z_{\rm f}$ of the viable region for inflation and dark matter for $C=10$ (left panel) and $C=30$ (right panel).  
The light gray region (region below the dotted line) is excluded by the SDSS (Ly$-\a$) data.
}
\label{fig:zc}
\end{figure}

In Fig.~\ref{fig:zc} we show the redshift parameter at the transition, $z_{\rm f}$, as a function of the ALP mass, $m_\phi$,
for those points in the groups B and C satisfying the CAST and HB star bounds. 
Since $z_{\rm f}$ is independent of the ALP-photon coupling, the distribution is identical for the groups B and C. 
One finds that the suppression at small scales is consistent with the SDSS (Ly-$\alpha$) data for $m_\f\gtrsim 0.01 (0.05)\,\EV$.

\vspace{5mm}

Interestingly, there are some anomalies which can be interpreted as a hint for the ALPs in 
the ALP miracle region.
In a study of the ratio of the number of stars in the HB to the number of red giant branch 
in globular clusters, the authors of Refs.~\cite{Straniero:2015nvc,Ayala:2014pea} found
a preference for the existence of an ALP with
$g_{\phi \gamma \gamma} = (0.29 \pm 0.18) \times 10^{-10} {\rm\,GeV}^{-1}$.
Also, it was shown in Ref.~\cite{Agarwal:2014qca} 
 that the suppression of the small-scale structure in the late-forming dark matter scenario could 
lead to a better agreement with the observed number of the dwarf galaxies and relax 
the missing satellite problem~\cite{Klypin:1999uc, Moore:1999nt} if
\begin{equation}
\laq{missat}
z_{\rm f} = 1.5 \times 10^6
\end{equation}
or slightly smaller.
This is consistent with the ALP miracle region (see \Eq{zc} and  Fig.~\ref{fig:zc}).

Lastly let us mention a case in which the ALP is coupled to gluons.
Since the dissipation rate is independent of the gauge couplings (cf. \Eq{disEW}),
 the dissipation effect induced by the ALP-gluon coupling does not differ significantly from the weak bosons. 
 However, the supernova constraint on the gluon coupling is much tighter\cite{Raffelt:2006cw}, 
 \begin{equation} 
 \laq{glc}
 f > 4 \times 10^8{\rm\,GeV},
 \end{equation}
 where we assume a coupling to gluons of the form 
 \begin{equation}
 \frac{\alpha_s }{8 \pi} \frac{\phi}{f} {G_{a\mu\nu} \tl{G}^{a\mu \nu}},
 \end{equation}
 where $G_{a\mu\nu}$ and  $\tl{G}^{a\mu \nu}$ are the field strength of gluons and its dual, respectively. 
 For $f$ satisfying \Eq{glc}, the dissipation process would be too inefficient to realize $\xi \ll 1$. 
 Hence, we do not consider this case further.

\section{Reheating through couplings to fermions}
\lac{matter}
In this section we consider an interaction of the ALP with a SM fermion $\psi$ to see how the preferred parameter region is modified.
The Lagrangian is given by
\begin{equation}
\laq{int1}
{\cal L}_{\rm int}=c_\psi y_\psi \frac{i \f }{f} \( \p_{L} H \ol{\p}_R \)+h.c.
\end{equation}
where $c_\psi$ is a constant of order unity, and $H$, $\p_{L}$, and 
$\ol{\p}_{R}$ denote the Higgs field, a left-handed fermion, and a right-handed anti-fermion,
respectively. Here the Higgs field and the left-handed fermion form a doublet under SU(2)$_L$,
but the gauge and flavor indices are omitted for a concise notation. 
This interaction is obtained if the Yukawa coupling $y_\psi$ is interpreted as a spurion field 
charged under a $\U(1)$ symmetry which is spontaneously broken at a scale $f$.  
The corresponding pseudo Nambu-Goldstone boson is
identified with the ALP $\phi$.  In this case, $c_\psi$ corresponds to the charge of $y_\psi$ under the U(1) symmetry.

When the Higgs field develops a nonzero expectation value, $v = \langle H^0 \rangle$, the fermion $\psi$ acquires a mass,
$m_\psi=y_\psi v$. At a scale below $m_\psi$, the fermion can be integrated out, leaving
a coupling of the ALP with gauge fields through chiral anomaly. Specifically, the coupling to photons 
is given by
\begin{equation}
g_{\f\g\g} \equiv  \frac{n_c c_\psi q_\psi^2 \alpha   }{\pi f},
\label{gfgg}
\end{equation}
where $n_c$ is equal to $3(1)$ for a quark (a lepton) and $q_\psi$ denotes the charge of the fermion. Similarly, if $\psi$ is a quark,  the ALP-gluon coupling is induced.

 \begin{figure}[t!]
\begin{center}  
   \includegraphics[width=105mm]{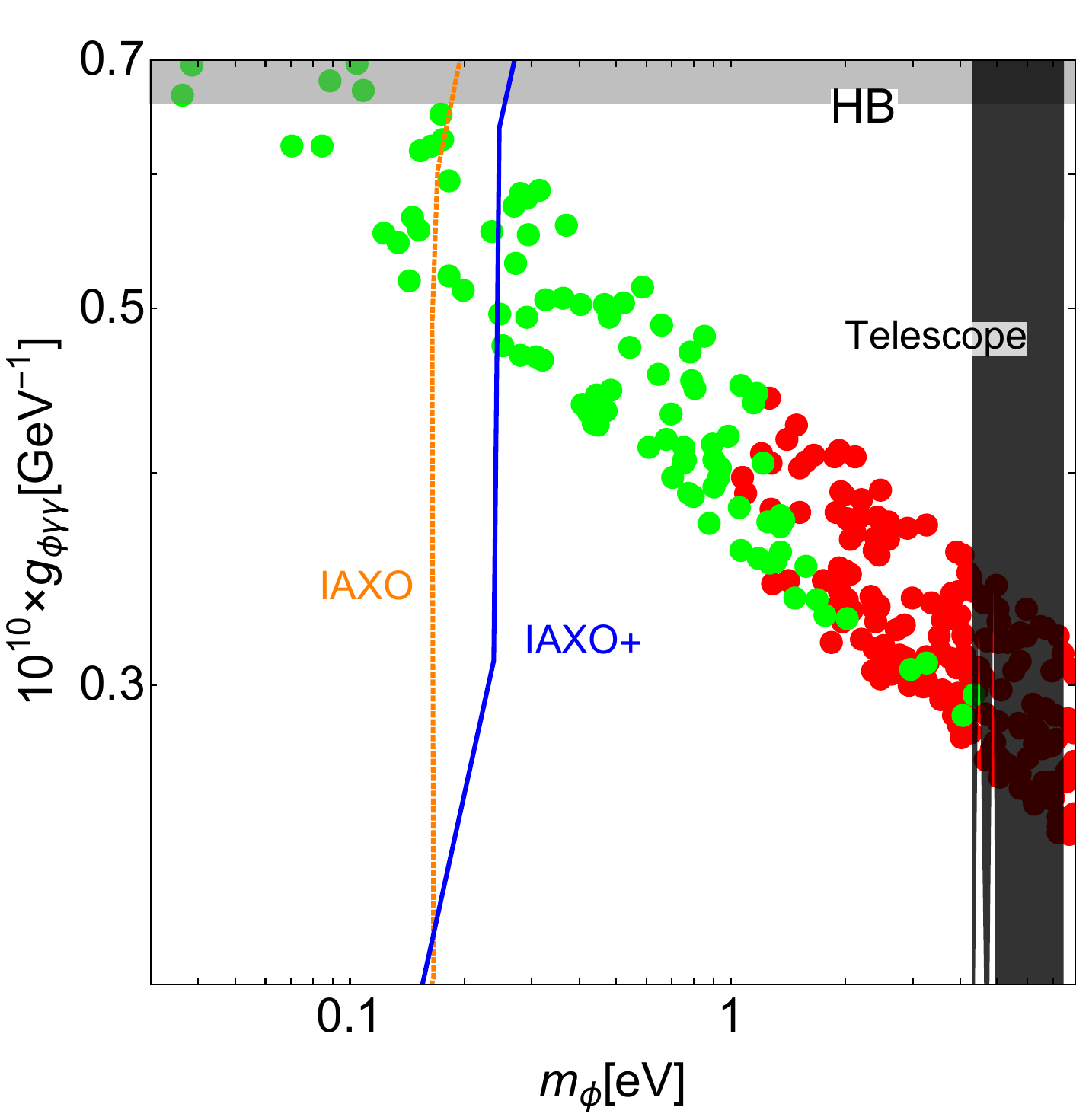}
\end{center}
\caption{
 Scatter plot of the viable parameters consistent with the CMB observation and the dark matter abundance, if the reheating
 proceeds mainly through the ALP coupling with $\tau$ leptons.  
 The red (green) points represent that $c_\tau$ is greater (smaller) than $1$. }
\label{fig:matDM}
\end{figure}

In this section we neglect the evaporation through such couplings to gauge fields induced by
the SM fermion loop, because the dissipation effect is suppressed at $T <m_\psi$ for the
parameters of our interest. Instead, 
the reheating of the ALP condensate mainly proceeds through the matter coupling \eq{int1}.
The subsequent cosmological evolution after reheating is basically same as in the previous section.

After inflation, the ALP condensate first decays into the matter fields. 
It decays into two fermions,  $\f \rightarrow \psi + \bar\psi$,
with a rate given by
\begin{equation}
\laq{decf}
\G_{{\rm dec},\psi}\simeq
\frac{n_c }{8\pi } \(\frac{c_\psi m_\psi }{f}\)^2m_{\rm eff}.
\end{equation}
The produced fermions quickly form thermal plasma. When the thermal mass exceeds the effective mass of the ALP
condensate,  $eT\gtrsim m_{\rm eff}$, the perturbative decay stops due to the thermal effect. Afterward, the dissipation process 
such as $\phi+ \psi \rightarrow \psi + \gamma \,({\rm or~} g)$ becomes significant,  and its  rate is given by \cite{Mukaida:2012bz}
\begin{equation}
\laq{disbrF}
\G_{{\rm dis},\psi}\simeq \displaystyle{ A_0 n_c\(\frac{c_\psi m_\psi  }{f}\)^2 \(\frac{ \a_\psi}{2\pi^2}\) T,}
\end{equation} 
where $A_0$ is a numerical constant. In the following we set $A_0=0.5$, and $\a_\psi=\a$ for leptons and $\a_\psi=\a_s$ for quarks.
When the temperature becomes higher than $T_{\rm EW}$, the Higgs boson is thermalized and the dissipation process such as
$\phi+{\rm Higgs} \rightarrow \psi + \bar \psi$ takes place.
The dissipation rate of the process involving the Higgs field in the initial or final state is given by\cite{Salvio:2013iaa} 
\begin{equation}
\laq{disF}
\G_{{\rm dis},\psi H}\simeq  
n_c  \(\frac{c_\psi^2 y_\psi^2  }{2 \pi^3 f^2}\) T^3, 
\end{equation} 
which is valid for $T > T_{\rm EW}$.

For successful reheating, the dissipation rate should be greater than or comparable to 
the Hubble parameter even when most of the inflaton energy 
already turns into radiation.
Assuming that the temperature is higher than the electroweak scale, the condition reads
\begin{equation}
\G_{{\rm dis}, \psi H}\gtrsim H
\simeq \sqrt{\frac{\pi^2 g_*  }{90}} \frac{ T^2}{M_{pl}}, 
\laq{recon2}
\end{equation}
where we have approximated that the Universe is radiation dominated in the second equality. In fact, 
the inequality should be approximately saturated in order for the relic ALP condensate to explain dark matter,
since otherwise the relic ALP abundance would be too small. 
Then,  using $V_0 \simeq 48 \lambda f^4/n^2$, one obtains
\begin{equation}
\laq{phigg}
g_{\f\g\g} 
\sim 3\times 10^{-12} 
\GEV^{-1} \(\frac{q_\psi^2 }{c_\psi}\) \(\frac{ 0.01 }{{y_\psi}}\)^2 \(\frac{10^{-12} }{\l}\)^{\frac{1}{4}},
\end{equation}
for successful reheating and explanation of dark matter.
Comparing \eq{phigg} with the current bound on $g_{\phi \gamma \gamma}$ in \eq{cast}, one finds that the 
couplings with light fermions with $y_\psi \lesssim 10^{-3}$ do not lead to successful reheating. 
In a case of the coupling to leptons,  only the coupling with $\tau$ leads to successful reheating. 
In a case of the coupling to quarks, the required decay constant becomes larger because of the larger 
dissipation rate. However, as we shall see in the next section, there is
an upper bound on the ALP mass as well as on the decay constant  (see \Eq{HDc}), because
thermally produced ALPs contribute to a hot dark matter component. In the case of a top quark, it is difficult
to satisfy the hot dark matter constraint unless a very small $c_\psi$ of ${\cal O}(0.01)$ is 
assumed.\footnote{If  the ALP is not the main component of dark matter, one may consider a simple case in which
the ALP couples to all the fermions with $c_\psi=\O(0.1)$ and $f= \O(10^8)\GEV$.
From \Eq{phigg} one finds that the ALP condensate completely evaporates due to the ALP-top interaction,
leaving only negligible amount of the ALP condensate.  The coupling to electrons in this case
is consistent with the excessive cooling of the white dwarf and the red giant stars\cite{Giannotti:2015kwo}.
} 
In the case of a bottom quark, the constraints, \eq{glc}
and from the optical telescopes can be only marginally satisfied.
In the following we therefore focus on the coupling with $\tau$.\footnote{
If there is a flavor violating interaction with $\mu$-lepton as $\frac{c }{f} \ol{\p}_{\tau L} \partial^\m\f  \g_\m {\p}_{\m L}$, 
the flavor violating decay of $\tau^- \rightarrow \m^- + {\rm ALP}$ takes place with a rate that may be tested by the  Belle II experiment
depending on the value of $c$. (cf. \cite{Yoshinobu:2017jti}.)
}

Now let us estimate the abundance of the remnant ALP condensate in the case of $\psi=\tau$
by solving the Boltzmann equations Eqs.(\ref{evolution}) with
\begin{align}
\G_{\rm tot}=\left\{
\begin{array}{ll}
{\displaystyle{ \G_{{\rm dec,}\psi}+\G_{{\rm dis,}\psi}~ (T<T_{\rm EW})}}\\  &\\ 
\displaystyle{
\G_{{\rm dis}, H\psi}~(T>T_{\rm EW})} 
\end{array} \right..
\end{align}
When the perturbative decay of $\f\rightarrow \t +\bar \t + {\rm Higgs}$ is kinematically allowed, this decay rate is also 
taken into account.

To identify the viable parameters, we first take about $100$ points in the viable 
inflaton parameters in the range of $-0.04 <(\kappa-1)\(f/M_{\rm pl}\)^{-2}<0.1$ and $0 <\h  \(f/M_{\rm pl}\)^{-3}<0.08$ for a given decay constant $f$ (cf. Fig.~\ref{fig:ns}). We vary the decay constant 
 from 
$f=5\times 10^6\GEV$ to $2\times 10^8\GEV$ by an interval of $5 \times 10^6\GEV$.
For each point in the viable region, we randomly generated  $20$ values of $c_\tau$  in the range of $0<c_\tau <5$.
Then we estimate the relic ALP abundance, $\Omega_{\f}h^2$, by solving Eqs.~(\ref{evolution}) for each set of parameters, $(c_\tau, f, m_\f, V_0)$. 
In Fig.~\ref{fig:matDM} we show those points satisfying $0.08<\Omega_{\f} h^2<0.16$.
The range of $c_\tau$  is found to be between $0.4$ and $1.7$.
 The slope of the scattered points can be understood by noting that 
 $g_{\f\g\g} \propto 1/c_\tau$ and $f \propto c_\tau^2$ from (\ref{gfgg}) and \eq{phigg},
 leading to $m_\f \sim \frac{\sqrt{\l} f^2 }{M{pl}} \propto c_\tau^4 \propto g_{\f\g\g}^{-4}$. 
In contrast to the previous case,  most of the viable parameters extend beyond the reach of the future solar axion search experiments.
The decay of the ALP dark matter with $m_\phi \gtrsim 3\,\EV$ may contribute to
 the diffuse cosmic infrared background spectrum \cite{Gong:2015hke,Kohri:2017oqn}.  
The region may also be tested in the photon-photon collider.

\section{Thermalized ALPs as hot dark matter}
\lac{HD}

One of the robust predictions of our scenario is the existence of thermalized components of the ALPs.
While most of the energy of the ALP condensate evaporates into thermal plasma,
 the ALPs are also thermally produced in the plasma.  The produced ALPs decouples at $T = T_{\rm d}\gtrsim T_{\rm EW}$, if one considers couplings to the weak gauge bosons or tau lepton~\cite{Salvio:2013iaa}. Then the abundance of thermalized ALPs is expressed as
\begin{equation}
\Delta N_{\rm eff}\simeq0.027\left(\frac{106.75}{g_{*,{\rm vis}}(T_{\rm d})}\right)^{4/3}
\end{equation}
in terms of the effective number of extra neutrino. Here $g_{*,{\rm vis}}(T)$ is the effective degrees of freedom of the SM plasma.
The typical prediction $\D{N}_{\rm eff}\sim 0.03$ can be tested by the future observations of CMB and 
BAO~\cite{Kogut:2011xw, Abazajian:2016yjj,Baumann:2017lmt}.

After the electron-positron annihilation, the temperature of the thermalized ALPs is given by
\begin{equation}
T_{\phi}\simeq0.33\left(\frac{106.75}{g_{*,{\rm vis}}(T_{\rm d})}\right)^{1/3}T_{\gamma},
\end{equation}
where $T_{\gamma}$ is the photon temperature, where we have used the entropy conservation of the SM plasma. 
Since the typical ALP mass is of ${\cal O}(0.01-1)$ eV, the thermalized ALPs become non-relativistic around the recombination, and they
contribute to hot dark matter.
The CMB lensing and cosmic shear, combined with current CMB and BAO observations, set an upper bound on the mass 
of the ALP hot dark matter as
\begin{equation}
\laq{HDc}
m_{\f}< m_{\rm \f,HDM}^{\rm bound}\simeq 7.7 \,\EV \(\frac{0.03 }{\D N_{\rm eff} }\)^{3/4},
\end{equation}
where we have derived the upper bound on the ALP mass $m_{\rm \f,HDM}^{\rm bound}$ by translating the bound on the gravitino mass of Ref.~\cite{Osato:2016ixc}. 
Notice that this bound as well as the prediction of $\D N_{\rm eff}\sim 0.03$ is independent 
of whether the remnant ALP condensate becomes the dominant dark matter.

\section{Discussion and conclusions}
\lac{6}

So far we have studied the case in which the ALP is coupled to the SM fields. If the ALP is coupled to extra
vector-like matters, the evaporation of the ALP condensate may mainly proceed through the coupling.
For successful reheating, the extra matter must be relatively light and it can be searched for at collider experiments.
The mass of the extra matter must be lower than the reheating temperature,
\beq
T_R \lesssim 80 \TEV\(\frac{107.75}{g_*(T_R)} \)^{1/4} \(\frac{m_{\rm \f,HDM}^{\rm bound}}{8\EV }\)^{1/2},
\eeq
in order to be thermally populated. If the other couplings are neglected, the ALP condensate must decay into the extra matter fields, and so, the mass of the extra matter must be lower than half the maximum effective ALP mass,
\begin{equation}
\laq{effmax}
m_{\rm eff}^{\rm max} \sim 1 {\rm \,TeV} \(\frac{m_{\rm \f,HDM}^{\rm bound}}{8\,\EV }\)^{1/2} \(\frac{\l }{7\times 10^{-12}}\)^{1/4},
\end{equation}
where we have taken $\r=V_0$ in \Eq{eff} to evaluate the maximum value.\footnote{The precise value of the largest effective ALP mass after inflation depends on the tachyonic preheating as well as the 
inflaton parameters, and the right-hand side of \eq{effmax} contains an uncertainty of ${\cal O}(1)$.
}
Also the extra matter fields should have sizable couplings to the SM fields so that the SM particles are thermalized
and the frequent scattering with thermal plasma leads the dissipation rate like \Eq{disbrF}.

To be concrete, let us suppose that the ALP is coupled to a pair of charged vector-like leptons, 
a right-handed lepton $E_R$ and its partner $\ol{E}_R$, as
\begin{equation} 
\laq{intad} 
 {\mathcal \d L}=\frac{i c_{E} \f}{f}m_{E} E_R \ol{E}_R+h.c.,
\end{equation} 
where $c_E$ is the coupling constant, and $m_E$ is the mass of the vector-like lepton.
The mass of the additional leptons is constrained by the LEP as $m_{\rm E}\gtrsim 100\,\GEV$~\cite{Achard:2001qw}.
Thus, for successful reheating, it implies $m_{\rm eff}^{\rm max}\gtrsim 200\GEV$,
 or equivalently, $m_\f> \O(0.1)\,\EV$.
 We have numerically solved the Boltzmann equations with a dissipation rate given by \Eq{disbrF},
and found that there is a reasonably large parameter region within $100\GEV<m_{\rm E}<\O(1)\,\TEV$ and $c_E = \O(1)$
satisying all the constraints discussed so far. The extra lepton
$E_R$ may have a small mixing with a SM lepton. In this case it decays into the SM lepton plus a photon. 
If the extra lepton is sufficiently long-lived, it can be searched for at the high-luminosity LHC (cf. Refs.~\cite{llstau,llstau2,llstau3}). 
Also it can be searched for in the future lepton colliders such as CEPC, ILC, FCC-ee, and CLIC \cite{CEPC, CEPC2, Fujii:2015jha, dEnterria:2016sca, Abramowicz:2016zbo}.

In this paper, we have revisited the ALP miracle scenario~\cite{Daido:2017wwb}, where the ALP plays 
the role of both inflaton and dark matter in our Universe. First we have extended the previous analysis by
scanning the whole inflaton parameter space and delineated the viable parameter region consistent with the CMB observations.
Next we solved the Boltzmann equations by incorporating the uncertainties in the dissipation process as well as
the model-dependent couplings to the weak gauge bosons.
As a result we have found the viable (``ALP miracle'') region
as $0.01\,\EV\lesssim m_\phi \lesssim 1$\,eV and $g_{\phi \gamma \gamma} = {\cal O}(10^{-11})$\,GeV$^{-1}$.
Interestingly, the ALP miracle region is consistent with the recently reported cooling anomaly hinted by
the ratio of the HB stars to the red giants in globular clusters~\cite{Straniero:2015nvc,Ayala:2014pea}.
Furthermore, the relic ALP condensate starts to behave like dark matter at a late time,
which could ameliorate the missing satellite problem~\cite{Agarwal:2014qca}.

Finally we have discussed the reheating through the ALP coupling to SM fermions,
and we have found that the ALP with coupling to $\tau$ can lead to successful reheating, while satisfying various constraints. 
To account for the dark matter, the ALP mass should be $m_\phi =\O(0.1-1)$\,eV and 
correspondingly, the (chiral anomaly-induced) coupling to photons is 
 $g_{\phi \gamma \gamma} = {\cal O}(10^{-11})$\,GeV$^{-1}$.

In both scenarios, once most of the ALP condensate evaporates into plasma, 
the ALPs are thermalized during the reheating process and they behave as dark radiation in the early Universe, 
which contributes to the deviation of the effective neutrino number $\D N_{\rm eff}\sim 0.03$. Around the recombination
thermalized ALPs become hot dark matter, suppressing the small-scale structure. 
This may be tested by the future CMB and BAO observations.

\section*{Acknowledgments}
F.T. thanks M. Amin for fruitful discussion on the scalar resonance at the workshop ``Post-inflationary string cosmology" at University 
of Bologna and T. Sekiguchi for useful communication on the hot dark matter constraint on the gravitino mass. 
This work is supported by 
Tohoku University Division for Interdisciplinary Advanced Research and Education (R.D.),
 JSPS Research Fellowships for Young Scientists (R.D.),
JSPS KAKENHI Grant Numbers JP15H05889 (F.T.), JP15K21733 (F.T.), 
JP26247042 (F.T),  JP17H02875 (F.T.), JP17H02878 (F.T.), and
by World Premier International Research Center Initiative (WPI Initiative), MEXT, Japan (F.T.).

\end{document}